\documentclass[11pt,preprintnubmers]{revtex4}
\usepackage{graphicx}

\begin{document}
\title{Remembering Sergio Fubini\footnote{Fubini Memorial, CERN, Geneva, May 2005}}
\author{R. Jackiw}
\affiliation{Center for Theoretical Physics\\
Massachusetts Institute of Technology\\
Cambridge, MA 02139\\
{\small MIT-CTP-3628}}

\begin{abstract}
\vspace{2in}
\begin{center}Abstract\\  The author recollects Sergio Fubini's impact on field theory\\ (radial quantization, merons, conformal quantum mechanics) and on MIT.\end{center}
\end{abstract}

\maketitle

\newpage
\thispagestyle{empty}
\begin{center}
\includegraphics[width=7in]{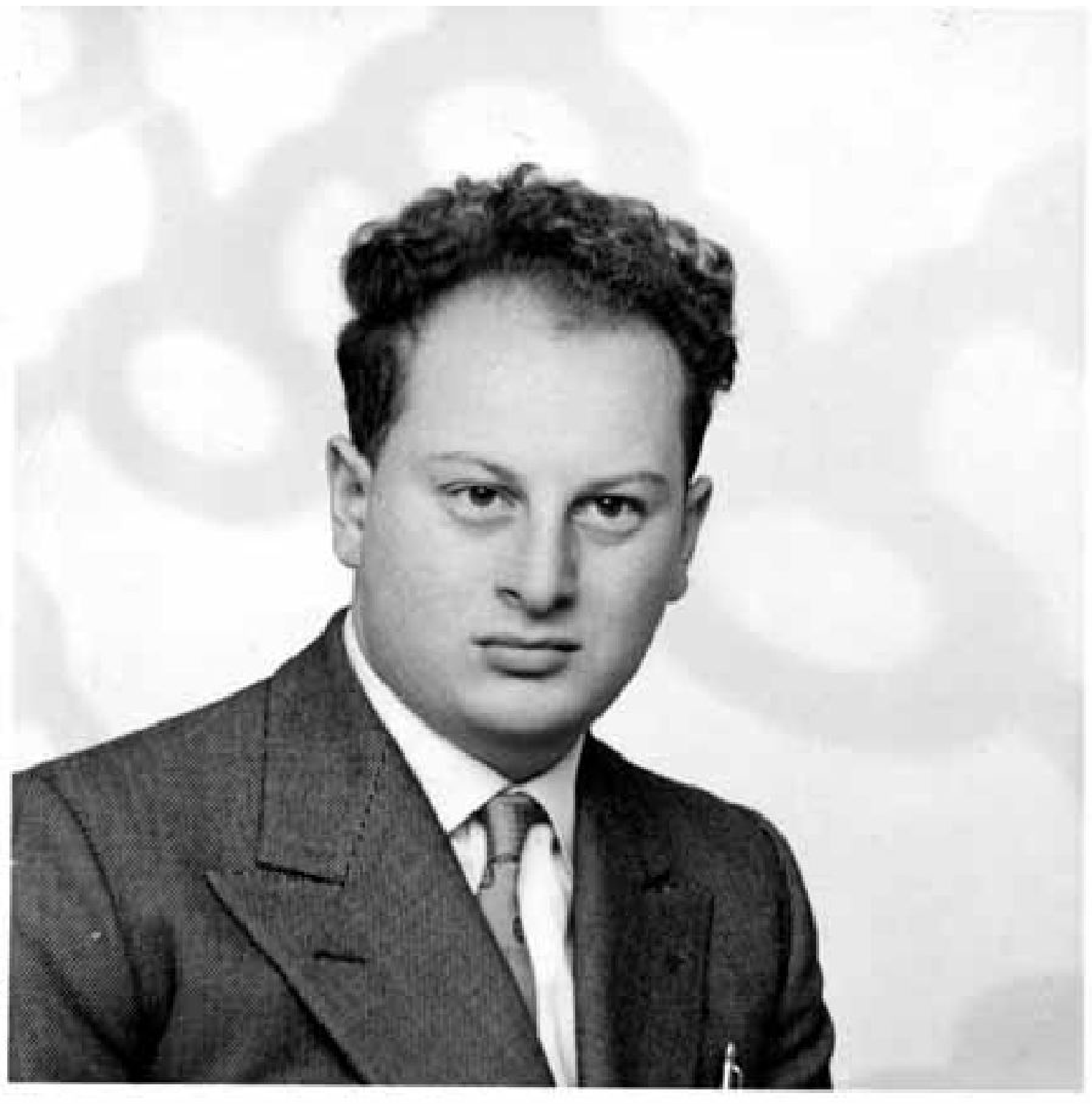} 
{\large \bf Sergio Fubini\\
                         1928-2005}
                         \vfill
\end{center}
\newpage

Sergio Fubini was in Boston in 1967, delivering the Loeb lectures at Harvard. I was a postdoc there and so I met him for the first time. In the same period Steven Weinberg also visited and lectured. It must have been clear that both were interested in staying in the area (at least for a while) because Viki Weisskopf, the very astute chairman of MIT's physics department, succeeded in hiring them -- a fortunate development that caused MIT to become the fountainhead of the principal themes in fundamental physics for the remainder of the 20th century. This was a direct consequence of the research that the two accomplished in their years at MIT: Weinberg reviving quantum field theory, unifying particle physics forces and discussing cosmology; Fubini developing an operator formalism based on the Veneziano amplitude, going beyond field-theoretic descriptions of Nature and leading to today's string theory.

Here we are commemorating Fubini, and Gabriele Veneziano already described their seminal string theory work at MIT. I shall speak about our joint work on field theory, and also about related investigations that Sergio subsequently carried out. 

Urged by Fubini and Weinberg, and wishing to be their colleague, I joined the MIT physics department in 1969. In my research I was finding new phenomena within current algebra and symmetry behavior in quantum field theory. These were subjects for which Fubini and his collaborators established fundamental results: approximate symmetries described by chiral currents; procedures for deriving physically useful sum rules from equal-time current commutators; the infinite momentum frame. \cite{fub1}

At that time, the findings by MIT-SLAC experimentalists of scaling in high energy (deep inelastic) electron scattering called attention to scale and conformal symmetries \cite{fub2}.  Moreover, the observed scattering amplitudes become determined at high energies by the behavior of relevant currents near the light cone \cite{fub3}. These observations led to the technique of ``light-cone" quantization, first discussed by Paul Dirac \cite{fub4}. In this approach, which is equivalent to the usual equal-time quantization, canonical commutation relations are posited on light-like surfaces, rather than at equal times. My collaborators and I took these ideas to the further step of deriving and postulating light-cone commutators for currents \cite{fub5}. This interested Sergio very much, because it turned out that the light-cone technique gives an operator formulation for his infinite momentum frame and results in a more efficient and accurate derivation of the sum rules, which he first discussed on the basis of equal-time current commutators.

I expect that this confluence of our research streams led to our collaboration. We took another quantization idea from Dirac, \cite{fub4} and developed a canonical  formalism for  fields by positing canonical commutation relations on surfaces of fixed $X^\mu X_\mu$,  with dynamical evolution proceeding in the direction normal to these surfaces \cite{fub6}. This manifestly Lorentz invariant formulation is especially convenient in Euclidean space, where fixed $X^\mu X_\mu$ defines a sphere, and evolution in the radial direction, normal to the sphere, is governed by the dilation generator $D$. Thus scale and conformally invariant theories fit very nicely into this approach, since there the dilation generator $D$ is a constant of motion (like the Hamiltonian $H$ in usual equal-time quantization of time-independent systems).
Initially our results appeared to be only of methodological, rather than practical, interest. Because in radial Euclidean quantization the  kinematical space at fixed radius is finite, we could give well-defined formulas for the Virasoro generators as moments of the energy-momentum tensor in 2-dimensional, conformally invariant theories. Moreover, the important center in that algebra was identified with the anomalous Schwinger terms in the ``equal-radius" commutators of the energy-momentum tensor. In this way another bridge appeared between our earlier researches: linking chiral anomalies on the one side with scale/conformal anomalies on the other. Both symmetries rely on absence of mass terms for the fields. But Nature's evident abhorrence of masslessness is encoded in quantum field theory by anomalies. 
\includegraphics{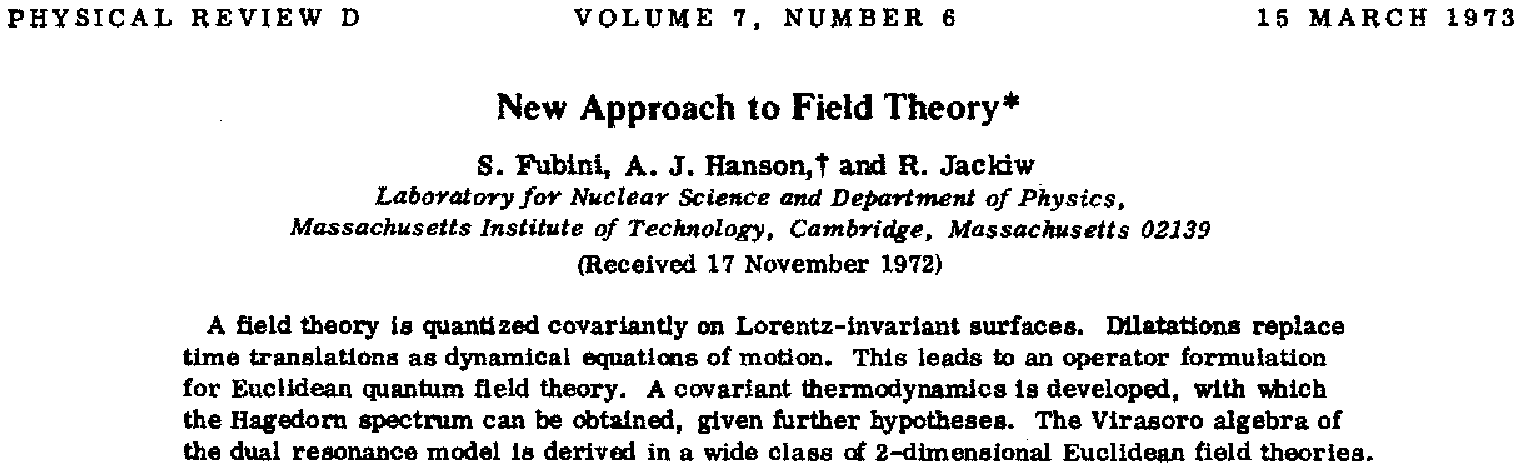}

Years later, when string theory revived, our radial quantization procedure became generally adopted as the preferred method for quantizing 2-dimensional, conformally symmetric models, which are at the core of string theory \cite{fub7}. 

After this joint work, Sergio left for CERN in 1973, and our direct collaboration ended. However, I was fortunate that a virtual collaboration continued, with both of us pursuing further topics of contemporaneous and mutual interest. The 1970's were a time for exploring field theoretic classical solutions and establishing their quantum meaning. The most important of these for physics was the Euclidean Yang-Mills instanton, found by a soviet of physicists \cite{fub8}.  The conformal invariance of Yang-Mills theory enlarges its ISO(3,1) Poincar\'e symmetry to an O(4,2) conformal symmetry, which becomes O(5,1) for the Euclidean theory. Claudio Rebbi and I showed that the instanton is O(5) invariant \cite{fub9}. Fubini, with Vittorio de Alfaro and Giuseppe Furlan, considered the Lorentzian theory and found regular solutions that preserve the O(4) $\times$ O(2) subgroup of O(4,2) \cite{fub10}. Thus both solutions are symmetric under the maximal compact group on their respective spaces. Moreover, in the Euclidean version, the Fubini {\it et al.} solution acquires localized singularities and in minimal form carries half of the instanton's topological quantum number. 

\includegraphics[scale=.90]{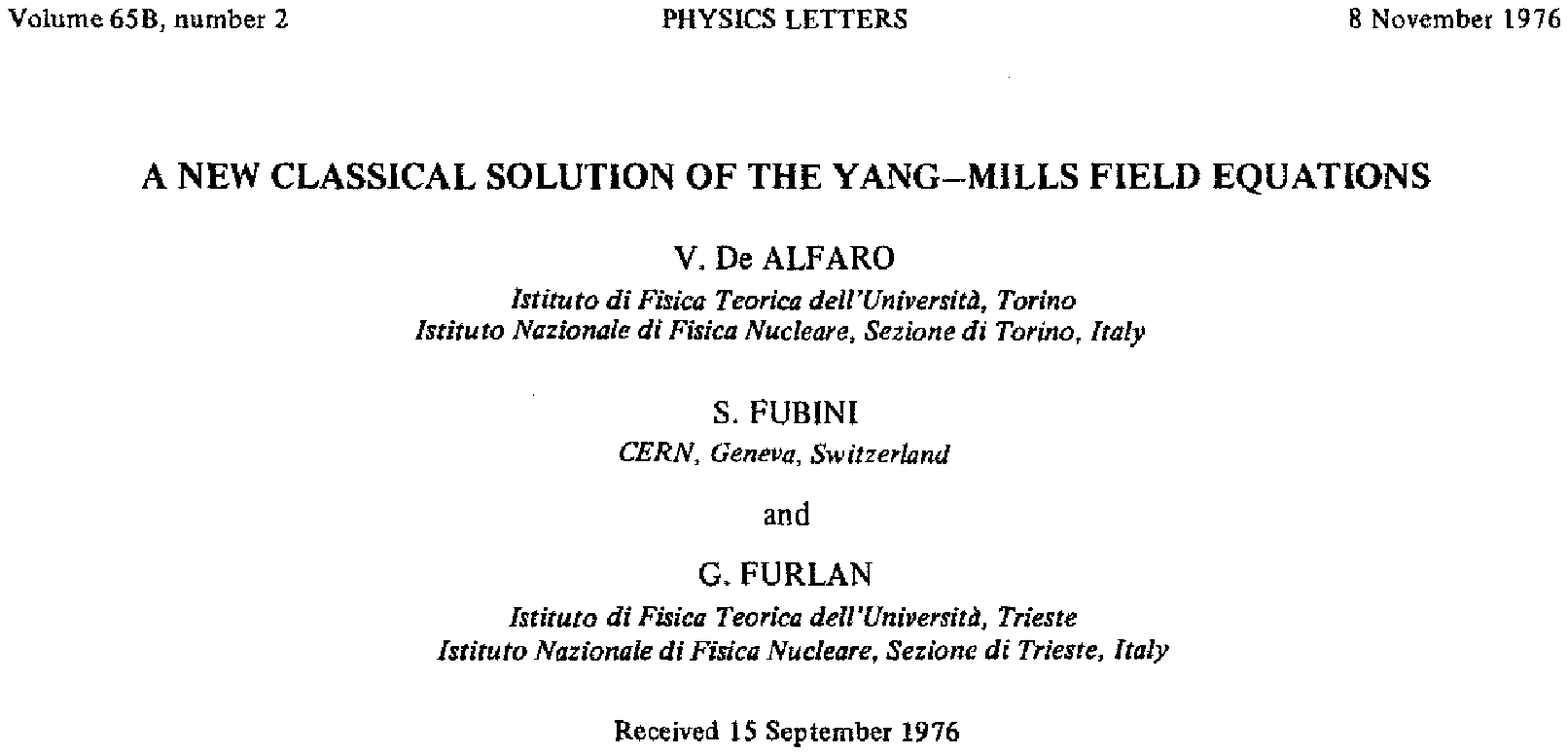}\\

\includegraphics[scale=.90]{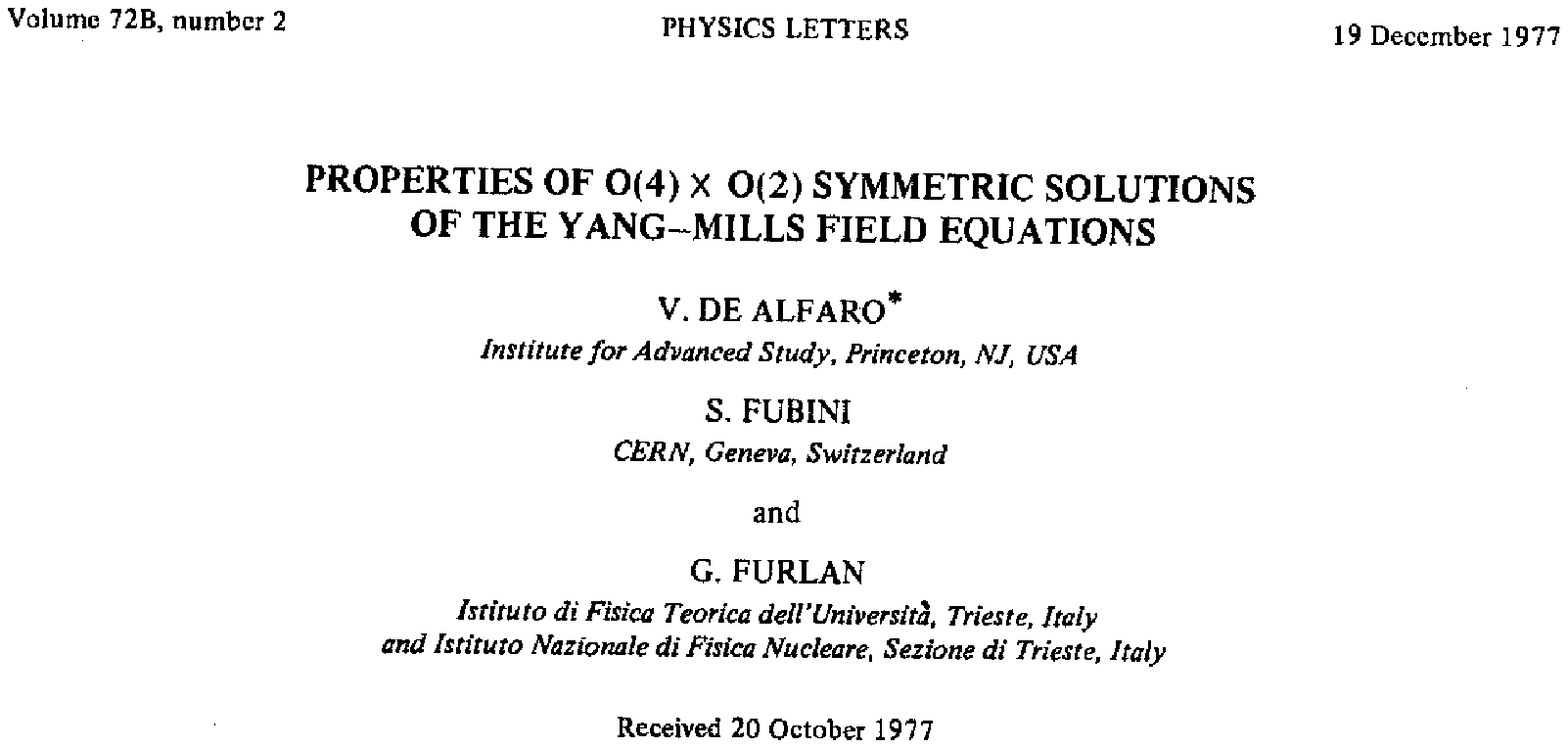}

The result lay fallow until the Princeton group \cite{fub11} suggested that the new solution should be called a ``meron" (after the Greek word for ``portion" [of an instanton]) and that merons cause color confinement. This idea has not been universally accepted, although it continues generating research to this day \cite{fub12}.

An alternative quantization in a conformally invariant model  was put forward by Fubini within particle quantum dynamics, which can be viewed as a ``field theory" in one time and zero space dimensions. When the dynamics is also conformally invariant, its time translation ``ISO(1,0)" symmetry becomes enlarged to O(2,1), with three generators. Compact rotations are generated by one of them, R: the time translation generator $H$ summed with the conformal generator $K, R \propto H + K$. [The third generator is the dilation $D$.] In the quantum theory one may choose to diagonolize $R$, which  has a discrete spectrum. This choice replaces conventional diagonalization of $H$ and/or radial quantization with its diagonalization of $D$. With de Alfaro and Furlan, Fubini applied this method to the quantum mechanics of the $1/r^2$ potential, known to be scale/conformal invariant \cite{fub2}. A very elegant group theoretical analysis ensued, and ``conformal quantum mechanics" was born \cite{fub13}.

\includegraphics{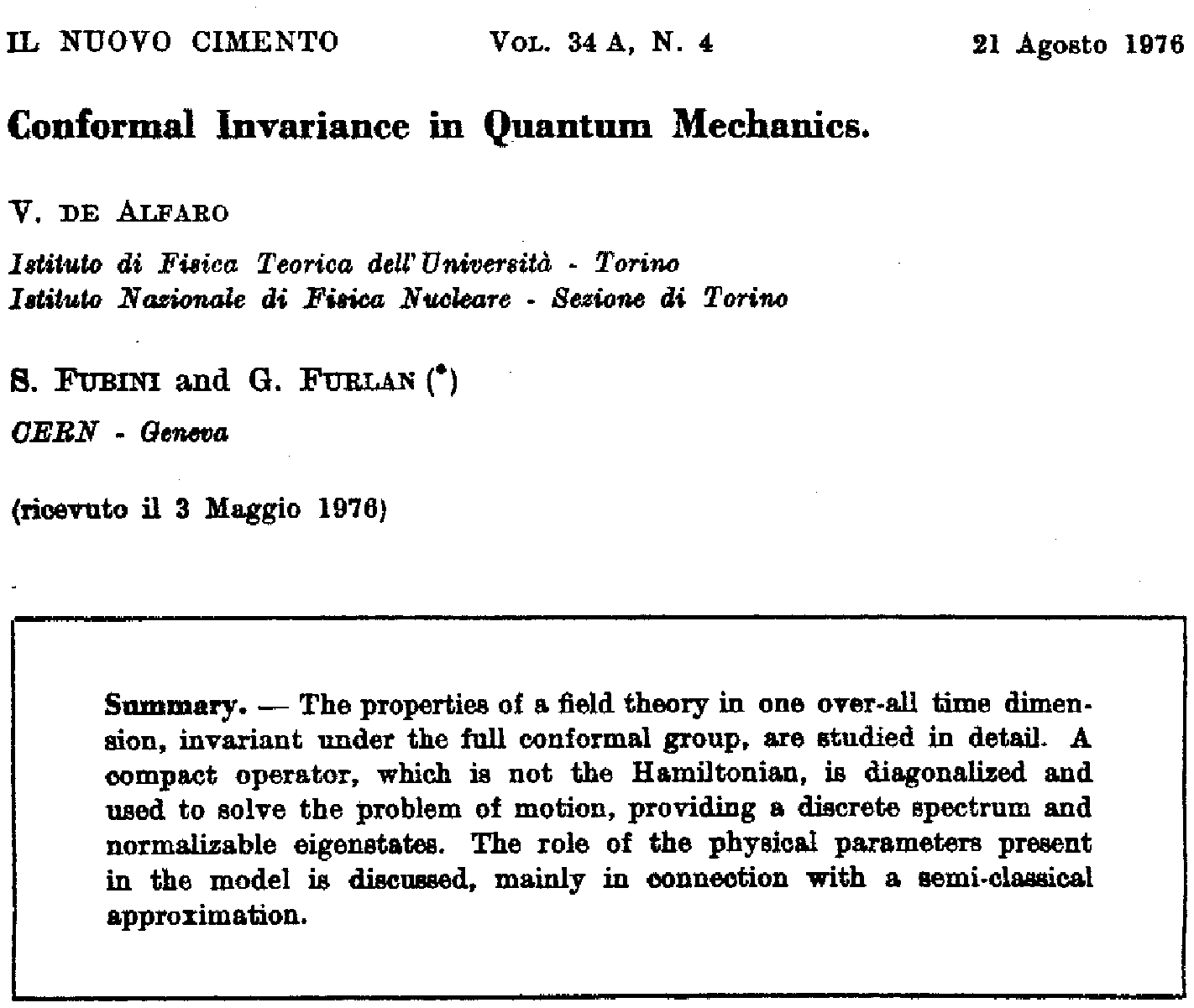}

But its utility at first remained obscure. It was established that interactions with magnetic point monopoles and vortices also preserve conformal symmetry and could fit into conformal quantum mechanics \cite{fub14}. Moreover, it was shown that the higher O(2,1) symmetry of the time-dependent Schr\"{o}dinger equation allows separating variables in ways other than the conventional time/space factorization. Thus conformal quantum mechanics arises from such an unconventional separation of variables in the O(2,1) invariant, time dependent Schr\"{o}dinger equation \cite{fub14}. [A familiar analogous situation arises with the 1/r potential, which in addition to rotational invariance, enjoys an O(4) symmetry; correspondingly the Schr\"{o}dinger equation can be separated in spherical and parabolic coordinates.] Fubini also demonstrated that similar conformal quantization can be carried out for conformally invariant field theories on space-time, but space-translation invariance is lost, to be replaced by a statistical conservation law for spatial momentum \cite{fub15}. Although this mechanism has not found favor, it anticipates contemporary interest in violations of space-time symmetries.

\includegraphics{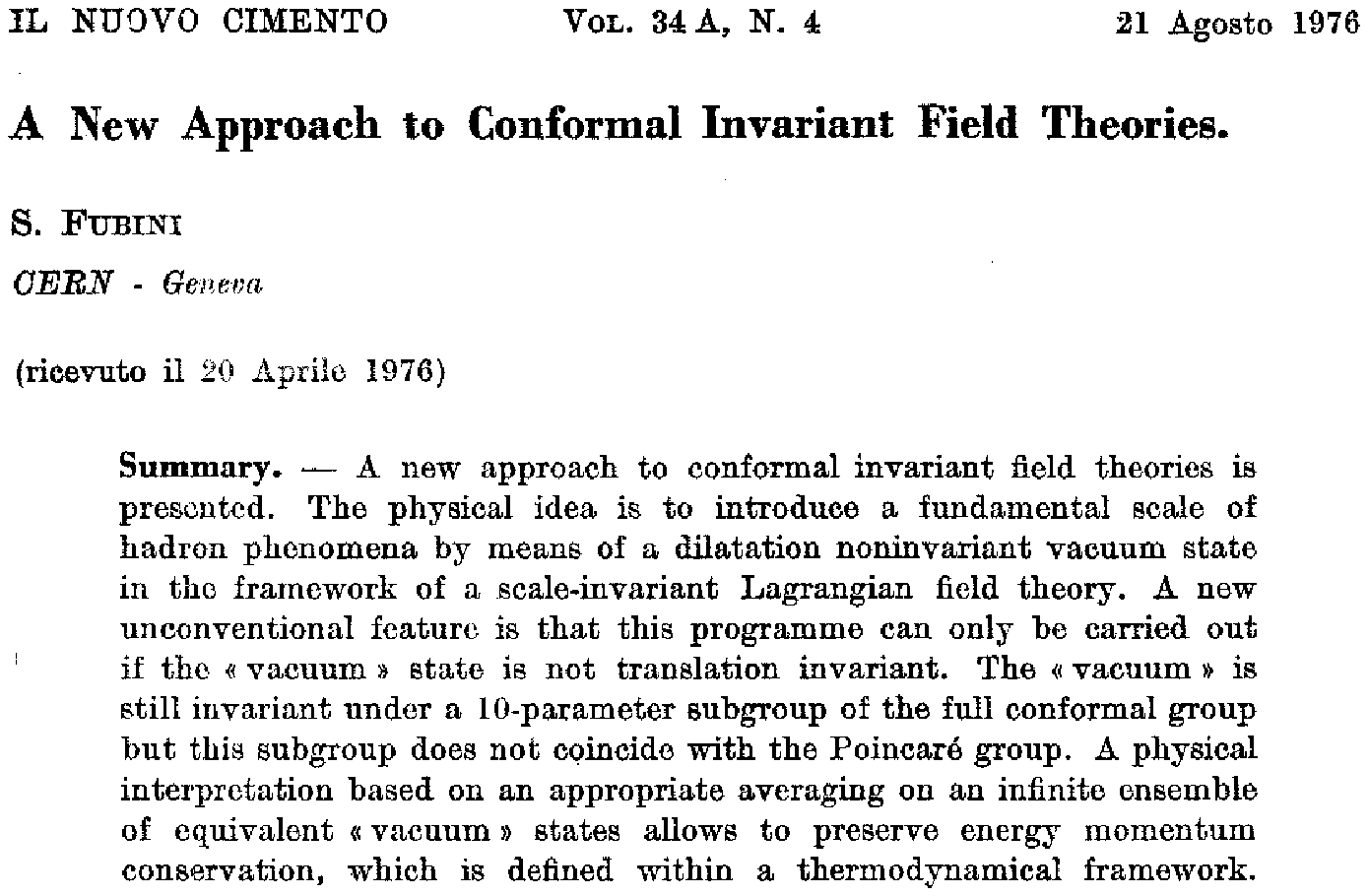}

The full impact of conformal quantum mechanics came only after string theorist realized that dynamics of a particle near a black hole may enjoy an O(2,1) symmetry, and the Fubini {\it et al.} choice of $R$ as the  evolution operator corresponds to a redefinition of the time variable near the black hole \cite{fub16}.

Thus it turns out that no matter how far Sergio's research strays from string theory, its final relevance reverts to string theory. Indeed an overview of his physics career shows that in early days he approached fearlessly dynamics of elementary particles, even where fundamental laws were unknown. This characterizes his work on the multiperipheral model, his sum rules and superconvergence relations, and of course the dual resonance model. But it seems to me that the mathematical depths that he plumbed in developing the latter into string theory awakened in him an interest in mathematical formalism, for which he surely had a genetic predisposition from his forbears. Nevertheless, Sergio's physical intuition prevailed and his explorations of formal mathematical physics, unified by the themes of conformal symmetry and alternative quantization procedures, found unanticipated applications.

 Fubini left MIT in 1973, but his impact on our physics department persisted, and continues to this day. He raised the visibility of the particle physics group by centering at MIT initial string theory research. Also one of the conditions of his employment was an agreement for an exchange program with Turin University. This brought to MIT countless Italian students, post-docs and senior visitors -- not only from Turin -- who were eagerly participating in the rich physics activity generated for the nascent string theory by Fubini and Veneziano, and also turning to other areas of fundamental physics. On the faculty level Fubini encouraged and supported his compatriots Rebbi and Veneziano on term appointments and Bruno Coppi in a permanent position at MIT.
 
 The flow of people continued informally after he left, but later it was formalized by a new INFN-MIT agreement, which established the ``Bruno Rossi" exchange program, named after an earlier Italian, who made his professional home at MIT and, like Sergio, did wonderful physics.

I last saw Sergio during a visit to CERN, where we met for lunch. During the meal, he became ill, so I insisted on driving him home to St. Cergue. He was very moved by my offer, and said that I treat him like a brother. I realize that for many years he was like an older brother, advising me professionally and encouraging me to enter new activities, like his Middle East peace initiative. That effort, similar to some of his physics researches,  awaits success only in the future, even while today it puts into vivid evidence his great talents in the service of humane pursuits.


\begin{thebibliography}{99}
\bibitem{fub1}
A guide to this research is V. de Alfaro, S. Fubini, G. Furlan and C.  Rossetti, {\it Currents in Hadron Physics} (North Holland, Amsterdam, 1973).

\bibitem{fub2}
R. Jackiw, ``Introducing Scale Symmetry," Physics Today, {\bf 25} (1), 23 (1972).


\bibitem{fub3}
  R.~Jackiw, R.~Van Royen and G.~B.~West,
  ``Measuring Light Cone Singularities,''
  Phys.\ Rev.\ D {\bf 2}, 2473 (1970).
  
  \bibitem{fub4}
  P.A.M. Dirac, ``Forms of  Relativistic Dynamics," 
Rev. Mod. Phys. {\bf 21}, 392, (1949).

\bibitem{fub5}
  J.~M.~Cornwall and R.~Jackiw,
  ``Canonical Light Cone Commutators,''
  Phys.\ Rev.\ D {\bf 4}, 367 (1971);
  D.~A.~Dicus, R.~Jackiw and V.~L.~Teplitz,
  ``Tests of Light Cone Commutators: Fixed Mass Sum Rules,''
  Phys.\ Rev.\ D {\bf 4}, 1733 (1971).
  
  \bibitem{fub6}
  S.~Fubini, A.~J.~Hanson and R.~Jackiw,
  ``New Approach to Field Theory,''
  Phys.\ Rev.\ D {\bf 7}, 1732 (1973).

\bibitem{fub7}
For a review, see P. Ginsparg, ``Applied Conformal Field Theory," in Les Houches Summer School (1988), [hep-th/9108028].

\bibitem{fub8}
  A.~A.~Belavin, A.~M.~Polyakov, A.~S.~Shvarts and Y.~S.~Tyupkin,
  ``Pseudoparticle Solutions of the Yang-Mills Equations,''
  Phys.\ Lett.\ {\bf B} {\bf 59}, 85 (1975).
  
 \bibitem{fub9}
  R.~Jackiw and C.~Rebbi,
 ``Conformal Properties of a Yang-Mills Pseudoparticle,''
  Phys.\ Rev.\ D {\bf 14}, 517 (1976).
  
  \bibitem{fub10}
  V.~de Alfaro, S.~Fubini and G.~Furlan,
  ``A New Classical Solution of the Yang-Mills Field Equations,''
  Phys.\ Lett.\ B {\bf 65}, 163 (1976);
  ``Properties of O(4) x O(2) Symmetric Solutions of the Yang-Mills Field
  Equations,''
  Phys.\ Lett.\ {\bf B} {\bf 72}, 203 (1977).
  
  \bibitem{fub11}
  C.~G.~Callan, R.~F.~Dashen and D.~J.~Gross,
  ``A Mechanism for Quark Confinement,''
  Phys.\ Lett.\ {\bf B} {\bf 66}, 375 (1977);
  ``Toward a Theory of the Strong Interactions,''
  Phys.\ Rev.\ D {\bf 17}, 2717 (1978).

\bibitem{fub12}
F. Lenz, J. Negele and M. Thies,
``Confinement form Merons,"
Phys. Rev. D {\bf 69}, 074009 (2004).

\bibitem{fub13}
  V.~de Alfaro, S.~Fubini and G.~Furlan,
  ``Conformal Invariance in Quantum Mechanics,''
  Nuovo Cim.\ A {\bf 34}, 569 (1976).
  
  \bibitem{fub14}
  R.~Jackiw,
  ``Dynamical Symmetry of the Magnetic Monopole,''
  Annals Phys.\  {\bf 129}, 183 (1980);
  ``Dynamical Symmetry of the Magnetic Vortex,''
  Annals Phys.\  {\bf 201}, 83 (1990).

\bibitem{fub15}
  S.~Fubini,
  ``A New Approach to Conformal Invariant Field Theories,''
  Nuovo Cim.\ A {\bf 34}, 521 (1976).
 
 \bibitem{fub16}
 For a review, see R. Kallosh ``Black Holes and Quantum Mechanics" in {\it Novelties in String Theory},
 L. Brink and R. Marnelius editors (World Scientific, Singapore, 1999).


\end{thebibliography}
\end{document}